\documentclass{aipproc}
\usepackage{amsmath,amssymb,enumerate}
\layoutstyle{6s}

\newcommand{\e}[2][]{
\begin{equation}
#2 \label{#1}
\end{equation} }

\begin{document}
\title{A locally anisotropic geometrical model of space--time based on CMBR}

\author{Panagiotis G. Labropoulos}{
  address={Physics Department, University of Athens,
Panepistimiopolis Zografou 15784, Athens, Greece, \sl{plabrop@phys.uoa.gr}}}

\author{Panayiotis C. Stavrinos}{
  address={Mathematics Department, University of Athens,
Panepistimiopolis Zografou 15784, Athens, Greece, \sl{pstavrin@math.uoa.gr}}}

\begin{abstract}
We study a locally anisotropic model of space--time induced by the existence of inflationary scalar fields in the framework of a more general geometrical structure than the Riemannian one. In this model the observable anisotropy of the CMBR (WMAP) is represented by a tensor of anisotropy and it is included in the metric structure of space-time.
As well, some interesting special cases of spaces are considered.  
\end{abstract}

\maketitle

\section{Introduction}
In 1994, Bekenstein, during his study on the applicability of Finsler geometry to physics \cite{bek1,bek2}, considered the following generalization of a conformal transformation which respects the weak equivalence principle and causality, but is not, though, angle-preserving:
\e[disformal]{\tilde g_{\alpha \beta }  = e^{2\psi } \left[ {A\left( I \right)g_{\alpha \beta }  + L^2 B\left( I \right)\psi _{,\alpha } \psi _{,\beta } } \right]}
with the following properties:
\begin{center}
\begin{enumerate}
	\item $B\left( I \right) \leq 0$
	\item $\begin{array}{*{20}c}
   {A\left( I \right) > 0,} & {C\left( I \right) = A(I) + IB(I)> 0}  \\ref

 \end{array} $
	\item $\frac{d}
{{dI}}\left[ {\frac{I}
{{C\left( I \right)}}} \right] > 0$
\label{disformal1}
\end{enumerate}
\end{center}
where $I$ is a invariant constructed by a scalar field $\psi$ and its first derivative $\psi _{\mu} $, $A, B, C$ are functions of $I$ and $L$ is a length scale. $C$ can be interpreted as a modulus of anisotropy. The meaning of the transformation (\ref{disformal}) is clear if it is examined under the prism of the Finslerian angular metric tensor: this transformation has an extra contribution which reaches its maximum along the gradient of the field.  Later, in 2003, Kaloper \cite{kaloper} studied an inflationary model with the gravitational coupling to matter being introduced by
\begin{equation}
\tilde g_{\mu \nu } \left( {x^{\mu}, \psi_{\mu} \left( x \right)} \right) = g_{\mu\nu} - \frac{1}{m^4}
\partial_\mu \psi \partial_\nu \psi \, .
\label{metrics}
\end{equation}
which is a special case of Bekenstein' s disformal transformation. The pseudo-scalar field $\psi$ here plays the role of the inflaton, which must describe a spin--0 particle to guarantee that inflation' s negative pressure condition is meaningful and has shift $\psi  \leftrightarrow \psi  + C$ and reflection $\psi  \leftrightarrow  - \psi $ symmetries to protect the inflaton from matter loop corrections. 

In this work we investigate some aspects induced by this model, by exploiting the rich geometrical structure of Finsler geometry and the non--linear connection on N--anholonomic manifolds in the framework of a vector bundle.

\section{Modified connection structure}
Let us consider the tangent bundle $TE$ of the total space $E$ of a fiber bundle. The study of this space is justified because if we would like to introduce a coordinate system on $E$ we need to define vectors over $E$, which belong to its tangent space. The fact that $E$ is constituted of fibers suggests a physical way to analyze a vector $X$ at a point $p$: $X$ must have a component tangent with respect to the fiber corresponding to $p$. So we can split $T_pE$ into two subspaces, one constructed of vectors tangent to the corresponding fiber called vertical $V_pE$, and its complimentary constructed of all the other vectors which belong to $T_pE$, called the horizontal subspace $H_pE$. This global splitting is represented by the Whitney sum of the non--intersecting subspaces:
\e{
{\rm T}_p E = V_p E \oplus H_p E
}
A base of the vertical subspace will be of the form $\left\{ {\frac{\partial }{{\partial \varphi _i }}} \right\}$, as these vectors are tangent to the fiber. The definition of the horizontal subspace is more complex as the demand of being complimentary to $V_pE$ is not sufficient. The reason beyond this is that the construction of the bundle does not suggest the points were neighboring fibers are glued, although we know the neighborhoods of these points. Generally, the vectors of $H_pE$ are linear combinations of $\frac{\partial }{{\partial x^i }}$ and ${\frac{\partial }{{\partial \varphi _i }}}$, so we can define the following base:
\e{\left\{ {\frac{\partial }{{\partial x^i }} + N_{ij} \left( {x,\varphi } \right)\frac{\partial }{{\partial \varphi _j }}} \right\}}
We can equip the bundle with a structure directly related to the parallel displacement of the elements of the fiber in order to compute the functions $N_{ij}$. This structure will provide us a rule of connecting fibers. This supplementary structure must be a connection structure.

We first consider a pseudo-Riemannian space--time as the base manifold and the space spanned by $\partial _\mu  \varphi $ as the typical fiber. Next we examine the case where the previous bundle is the base manifold and the fiber is spanned by a second field $\partial _\mu  \sigma$. The existence of a scalar field in this framework entering the metric by means of the so-called disformal transformation changes the way vectors are parallelly transported.
 
First we define for simplicity the covector $\varphi _i   = \frac{{\partial \psi }}{{\partial x^i  }} $. We notice immediately that our independent variables will be $\left( {x^\mu  ,\varphi _i  } \right)$. The adapted frame is set as follows:
\e[frame]{
\left\{ {\begin{array}{*{20}c}
   {dx^A  \equiv \left( {dx^\mu  ,\delta \varphi _i  = d\varphi _i  - M_{\nu i} dx^\nu  } \right)}  \\
   {\frac{\partial }{{\partial x^A }} \equiv \left( {\frac{\delta }{{\delta x^\nu  }} = \frac{\partial }{{\partial x^\nu  }} + M_{\nu i} \frac{\partial }{{\partial \varphi _i }},\frac{\partial }{{\partial \varphi _i }}} \right)}  \\
\end{array}} \right.
}
Then the connection structure is given by:
\e{
\Gamma _{BC}^A  = \left\{ {L_{\nu \lambda }^\mu  ,L_{j\lambda }^i ,h_{kl} C_\nu ^{\mu l} ,h_{kl} C_j^{il} } \right\}
}
The following covariant derivatives can be defined:
\e{
\begin{array}{*{20}c}
   {V^\mu  |_\lambda   = \frac{{\delta V^\mu  }}{{\delta x^\lambda  }} + L_{\nu \lambda }^\mu  V^\nu  }  \\
   \begin{array}{l}
 V^\mu  ||_i  = \frac{{\partial V^\mu  }}{{\partial \varphi _i }} + C_\nu ^{\mu i} V^\nu   \\ 
  
 \end{array}  \\
\end{array}  }
The metrical structure is introduced by:
\e[metricgn]{
\tilde g\left( {x^\mu  ,\varphi _i } \right) = g_{\mu \nu } dx^\mu   \otimes dx^\nu   + h^{ij} \delta \varphi _i  \otimes \delta \varphi _j }
where $h^{ij}$ is the metric of the vertical space $V_{p}E$ \cite{miron1}.

Had we started with a non--linear connection, instead of defining a priori the adapted frame, and determining the adapted frame we would have ended up with an off--diagonal metric of the following form:
\e{\tilde g = \left[ {\begin{array}{*{20}c}
   {g_{\mu \nu }  + M_{\mu i} {\rm M}_{\nu j} h^{ij} } & {M_{jk} h^{ik} }  \\
   {M_{ik} h^{jk} } & {h^{ij} }  \\
\end{array}} \right]}
The previous canonical metric structure is constructed by a Sasaki type lift from the base manifold. A number of ansatzs with off--diagonal metrics were investigated in higher
dimensional gravity (see, for instance, the Salam and Strathee and the Percacci and
Randjbar--Daemi works \cite{03sal}) which showed that off--diagonal components
in higher dimensional metrics are equi\-valent to including $U(1),SU(2)$ and
$SU(3)$ gauge fields. We can write the metric in the form (\ref{metricgn}) only if there exists a coordinate transformation under which it can be diagonalized.

We shall remark that the non--linear connection can be physically interpreted as the interaction between the inflaton and the matter sector. The metric is written in block form which expresses the independence of the inflaton' s motion in its "internal" space from the motion in space--time. 

No one restricts us from considering more light scalar fields. In the curvaton scenario a primordial fluctuation of a second scalar field $\sigma$ generates the density fluctuations needed to to explain the hierarchical large scale structure of the Universe. Several fields such as the Affleck-Dine field, pseudo-Nambu-Goldstone bosons, cosmological moduli fields, flat direction in the minimally supersymmetric model (MSSM) etc. are candidates for generating curvature perturbations. In this case we also define $\xi _a   = \frac{{\partial \sigma }}{{\partial x^a  }} $ and the independent variables become $\left( {x^\mu  ,\varphi _i ,\xi _\alpha  } \right)$. The spatial structure becomes second order vector bundle--like and the adapted frame is set as follows:
\e[frame2]{
\left\{ {\begin{array}{*{20}c}
   {dx^A  \equiv \left( {dx^\mu  ,\delta \varphi _i  = d\varphi _i  - M_{\nu i} dx^\nu  ,\delta \xi _\alpha   = d\xi _\alpha   - P_{\nu \alpha } dx^\nu   + Q_\alpha ^i d\varphi _i } \right)}  \\
   {\frac{\partial }{{\partial x^A }} \equiv \left( {\frac{\delta }{{\delta x^\nu  }} = \frac{\partial }{{\partial x^\nu  }} - M_{\nu i} \frac{\partial }{{\partial \varphi _i }} + P_{\nu \alpha } \frac{\partial }{{\partial \xi _\alpha  }},\frac{\delta }{{\delta \varphi _i }} = \frac{\partial }{{\partial \varphi _i }} - Q_\alpha ^i \frac{\partial }{{\partial \xi _\alpha  }},\frac{\partial }{{\partial \xi _\alpha  }}} \right)}  \\
\end{array}} \right.}
where we introduce three kinds of non-linear connections $\left\{ {M_{\nu i} ,P_{\nu \alpha } ,Q_\alpha ^i } \right\}$ representing inflaton--matter, curvaton--matter and inflaton--curvaton interactions respectively. Therefore, the connection and metrical structures are given by:
\e{
\Gamma _{BC}^A  = \left( {L_{\nu \lambda }^\mu  ,L_{j\lambda }^i ,L_{b\lambda }^a , C_{k \nu} ^{\mu}, C_{kj}^{i}, C_{kb}^{a} , H_{c \nu} ^{\mu } ,H_{cj}^{i}, H_{cb}^{a} } \right)
}
and the metric sstructure of this space--time becomes
\e{
\tilde g = g_{\mu \nu } dx^\mu   \otimes dx^\nu   + h^{ij} \delta \varphi _i  \otimes \delta \varphi _j  + k^{ab} \delta \xi _a  \otimes \delta \xi _b 
}
where $k^{ab}$ is the metric of a Hamilton space of order 2 \cite{miron1}.

In the single scalar field case the curvture tensors are 
\[
\begin{array}{l}
 R_{MN\mu }^\mu   = \frac{{\delta L_{MN}^\mu  }}{{\delta x^\mu  }} - \frac{{\delta L_{M\mu }^\mu  }}{{\delta x^N }} + L_{K\mu }^\mu  L_{MN}^K  - L_{KN}^\mu  L_{M\mu }^K  \\ 
 R_{MNi}^i  = \frac{{\delta C_{iMN} }}{{\delta \varphi _i }} - \frac{{\delta C_{iM}^i }}{{\delta \varphi _{\rm N} }} + C_{Ki}^i C_{MN}^K  - C_{KN}^i C_{Mi}^K  \\ 
 \end{array}
\]
and the Ricci tensors and curvature scalar are respectively
\e{{\cal R} _{MN}  = \left\{ {R_{MN\mu }^\mu  ,R_{MNi}^i } \right\}}

\e{
{\cal R}  = \tilde g^{MN} {\cal R} _{MN}  = g_{\mu \nu } R^{\mu \nu }  + h_{ij} R^{ij}  = R + R^{\left( {\inf laton} \right)} 
}
Although R is the usual Riemannian curvature scalar constructed from $g_{\mu \nu}$, $\cal R$ constructed from $\tilde g$ will differ because of the internal variables $h_{ij}$. This could indicate a way to solve the flatness problem through its contribution to $\Omega_{\kappa}$ \cite{efstathiou}.

\section{Deviation of Geodesics}
The relation that characterizes isotropic points in a Finslerian space--time of constant curvature is analogous to the Riemannian one and has the form \e{L_{\mu \nu \kappa \lambda }  = K\left( {\tilde g_{\mu \kappa } \tilde g_{\nu \lambda }  - \tilde g_{\mu \lambda } \tilde g_{\nu \kappa } } \right)}
where $K$ is the constant curvature. If we set into correspondence the events on two null geodesics $C$ and $C^{\prime}$ in a space of constant curvature their deviation reduces to:
\e{\frac{d}{{ds}}\left( {n^i \varphi _i } \right) + \varepsilon Kn^i \varphi _i  = 0}
where $\varepsilon  =  \pm 1$, $s$ is the conventional affine parameter and $n^i$ the deviation vector. This equation describes the rate of change of the cross--section of a beam of light rays passing through an inhomogenous mass distribution. In this space the equation of geodesic deviation may be integrated and the deviations of space--like and time--like geodesics are given by:
\e{
{\begin{array}{*{20}c}
   {\begin{array}{*{20}c}
   {{\rm K} > 0,} & {n^i \varphi _i  = c_1 \sin s\left( {\varepsilon K} \right)^{{\raise0.5ex\hbox{$\scriptstyle 1$}
\kern-0.1em/\kern-0.15em
\lower0.25ex\hbox{$\scriptstyle 2$}}}  + }  \\
\end{array}c_2 \cos s\left( {\varepsilon K} \right)^{{\raise0.5ex\hbox{$\scriptstyle 1$}
\kern-0.1em/\kern-0.15em
\lower0.25ex\hbox{$\scriptstyle 2$}}} }  \\
   {\begin{array}{*{20}c}
   {{\rm K} = 0,} & {n^i \varphi _i  = c_1 s + c_2 }  \\
\end{array}}  \\
   {\begin{array}{*{20}c}
   {{\rm K} < 0,} & {n^i \varphi _i  = c_1 \sinh s\left( { - \varepsilon K} \right)^{{\raise0.5ex\hbox{$\scriptstyle 1$}
\kern-0.1em/\kern-0.15em
\lower0.25ex\hbox{$\scriptstyle 2$}}}  + c_2 \cosh s\left( { - \varepsilon K} \right)^{{\raise0.5ex\hbox{$\scriptstyle 1$}
\kern-0.1em/\kern-0.15em
\lower0.25ex\hbox{$\scriptstyle 2$}}} }  \\
\end{array}}  \\
\end{array}}
}

Null geodesics are of great importance in relativity because nearly all astronomical information cames to us optically. The deviation vector in this case takes the form $n^2  = as^2  + bs + c$

We can define the Hubble parameter as $\tilde H = \frac{\dot{ \tilde a}} {\tilde a}$ where $\tilde a = \tilde a \left( {v\left( s \right)} \right)$ and $\dot {\tilde a} = \frac{{d \dot {\tilde a}}}{{ds}} = \frac{{\partial \dot {\tilde a}}}{{\partial v^i }}\frac{{dv^i }}{{ds}}$ with $v^i (s)$ the unit vector tangent to the flow lines. The Hubble parameter is defined as \cite{lesvos} $\tilde H = \frac{1}{3}\tilde \Theta  = v_{|\mu }^\mu   - C_{\mu \lambda }^\mu  \dot v^\lambda $ and the anisotropic character of the Hubble parameter is revealed by the Cartan torsion tensor. The Cartan torsion tensor is defined as $
C_{\mu \nu \lambda }  = \frac{1}{2}\frac{{\partial \tilde g_{\mu \nu } \left( {x,\varphi } \right)}}{{\partial \varphi _\lambda  }}
$.
This tensor which depends only on the coupling constant and the field' s derivative is very important because its vanishing is a criterion that the space is Riemannian. The Cartan torsion tensor plays the role of the Christoffel symbols with respect to $\varphi$. We must also note that in this form Hubble' s parameter accounts for a locally corrected version due the contribution from Cartan' s tensor.

For a scalar field $\psi$ the energy density $\rho _\psi  $ and pressure $p _\psi  $ are
$ \rho _\psi   = \frac{1}{2}\left( {\frac{{d\psi }}{{ds}}} \right)^2  + V\left( \psi  \right)$ and 
 $p_\psi   = \frac{1}{2}\left( {\frac{{d\psi }}{{ds}}} \right)^2  - V\left( \psi  \right)$
where $V\left( \psi  \right)$, so that the Einstein equations can be written in the form:
\e{L^{\mu \nu } \varphi _\mu  \varphi _\nu   = {\rm K}\left( {\left( {\frac{{d\psi }}{{ds}}} \right)^2  - V\left( \psi  \right)} \right)}
We also get the following form for the Raychauduri equation in the spitit of \cite{lesvos} $
\frac{d \tilde \Theta^2}{ds} = - \frac{1}{3} \tilde \Theta - \tilde \sigma_{ik} \tilde \sigma^{ik}+ \tilde \omega_{ik} \tilde \omega^{ik} - {\rm K}\left( {\left( {\frac{{d\psi }}{{ds}}} \right)^2  - V\left( \psi  \right)} \right)$. The continuity equation becomes $\dot \rho _\psi   =  - 3\tilde H\left( {\left( {\frac{{d\psi }}{{ds}}} \right)^2 } \right)$, yielding the inflaton field equation $\frac{{d^2 \psi }}{{ds^2 }} + 3\tilde H\frac{{d\psi }}{{ds}} =  - \frac{{dV}}{{d\psi }}$. The potential V remains to be taken from a more fundamental theory. Once we have inserted the potential we can calculate the slow roll parameters $ \varepsilon \left( \psi  \right) = \frac{{m_{Planck} ^2 }}{2}\left( {\frac{1}{V}\frac{{dV}}{{d\psi }}} \right) $ and $ \eta \left( \psi  \right) = m_{Planck} ^2 \left( {\frac{1}{V}\frac{{d^2 V}}{{d\psi ^2 }}} \right) $
and calculate the logarithmic derivative $
\frac{{d\ln P_\psi  }}{{d\ln \left( {\tilde a \tilde H} \right)}}
$
in order to calculate the scalar spectral index.

\end{document}